\documentclass[pre,twocolumn]{revtex4}
\usepackage{epsfig}
\usepackage{bm}
\usepackage{amsmath}
\usepackage{hyperref}
\usepackage{breakurl}

\begin{document}

\title{Multiplicity of chaotic and turbulent regimes in Rayleigh-B\'{e}nard convection}

\author{A. Bershadskii}

\affiliation{
ICAR, P.O. Box 31155, Jerusalem 91000, Israel
}

\begin{abstract}

 Multiple chaotic and turbulent regimes in Rayleigh-B\'{e}nard  convection have been studied and classified from the onset of deterministic chaos to the fully developed turbulence using the distributed chaos approach supported by results of laboratory experiments and numerical simulations. It is shown that the regimes can replace each other depending on value of the Rayleigh number and location in the convection cell (in the bulk of the flow or in the boundary layers). The inertial-buoyancy and helicity based invariants play crucial role in the formation of the regimes. Transition from the classical to ultimate state has been also discussed in this context.

\end{abstract}

\maketitle

\section{Inroduction}

  Multiplicity of different regimes in the Rayleigh-B\'{e}nard convection makes this important for applications problem rather difficult for investigation and comprehension. The main reason for this multiplicity is the non-linear interaction of the velocity and temperature fields. It is typical for such situations to look to invariants as a key for solution, which can clarify the matter. For chaotic and turbulent regimes the adiabatic invariants are actually the main tool of theoretical investigations. After Bolgiano and Obukhov pioneering papers \cite{bol},\cite{ob} the Kolmogorov-like phenomenology was applied to stably stratified fluids and then it was applied also for the Rayleigh-B\'{e}nard convection in the Refs. \cite{pz},\cite{lvov},\cite{fl} (see also Refs. \cite{lx},\cite{ch},\cite{vkp} and references therein). However, for the velocity and temperature spectra obtained in the experiments and direct numerical simulations the inertial range is usually too short to make definite conclusions. The distributed chaos approach allows to make an extension of the inertial range for the Kolmogorov-Bolgiano-Obukhov invariant phenomenology and provides good agreement with experimental and numerical simulations data. Also an additional ideal invariant - the helicity-based Levich-Tsinober invariant \cite{lt} - is considered in present paper for the Rayleigh-B\'{e}nard convection. An additional turbulent regime dominated by this invariant in the bulk of the flow is studied and compared with the results of laboratory experiments and direct numerical simulations. 
  
  In the Section II of the paper appearance of the distributed chaos from the deterministic one is studied and compared with results of a direct numerical simulation of the Rayleigh-B\'{e}nard convection.  In the Sections III and IV the Kolmogorov-Bolgiano-Obukhov and the buoyancy-helical distributed chaos have been introduced, studied and compared with the results of laboratory experiments and direct numerical simulations. And, finally, in the Section V the transition from the so-called classical to ultimate state is considered in the above mentioned context.

\section{From deterministic to distributed chaos}

The Boussinesq approximation for the buoyancy-driven flows is (see, for instance, Ref. \cite{kcv})
$$
\frac{\partial {\bf u}}{\partial t} + ({\bf u} \cdot \nabla) {\bf u}  =  -\frac{\nabla p}{\rho_0} + \sigma g \theta {\bf e}_z + \nu \nabla^2 {\bf u}   \eqno{(1)}
$$
$$
\frac{\partial \theta}{\partial t} + ({\bf u} \cdot \nabla) \theta  =  S  \frac{\Delta}{H}e_z u_z + \kappa \nabla^2 \theta, \eqno{(2)}
$$
$$
\nabla \cdot \bf u =  0 \eqno{(3)}
$$
here $p$ and $\theta$  are the pressure and temperature fluctuation fields ($\theta = T-T_0 (z)$ with $T_0(z)$ as a conduction-state profile), ${\bf u}$ is the velocity field, ${\bf e}_z$ is the vertical unit vector, the vertical distance between the two layers is $H$ and the temperature difference between them is $\Delta$ (in the case of the Rayleigh-B\'{e}nard convection the $H$ is distance between the bottom and top plates), $\rho_0$ is the mean density of the fluid, $\nu$ is the viscosity and $\kappa$ is the thermal diffusivity, the gravity acceleration is $g$ and the thermal expansion coefficient is $\sigma$.  For the Rayleigh-B\'{e}nard convection $S=+1$ whereas for the stably stratified flows $S=-1$.\\

  In direct numerical simulation of the Rayleigh-B\'{e}nard convection, reported in Ref. \cite{pcfg}, dynamics of a global characteristic - normalized heat current through the fluid layer $N(t)$ (the Nusselt number) at constant temperature of the top and bottom surfaces of a  cylindrical domain, was studied at the onset of the thermal convection at Prandtl number $Pr= \nu/\kappa = 0.78$ and aspect ratio $\Gamma = 4.72$. \\
  
  Figure 1 shows (in the log-log scales) a broadband windowed power spectrum of the $N(t)$ fluctuations obtained at $\delta = (Ra - Ra_c)/Ra_c = 0.614$, where $Ra_c$ is the Rayleigh number at the convective threshold ($Ra = \sigma g \Delta H^3/\nu\kappa$). The spectral data were taken from Fig. 3 of the Ref. \cite{pcfg}. Observations performed in the Ref. \cite{pcfg} indicate that the spectrum shown in the Fig. 1 is dominated mainly by the nucleation of dislocation pairs.

  The dashed curve in the Fig. 1 indicates the exponential spectrum 
$$ 
E(f) \propto \exp(-f/f_c)  \eqno{(4)}
$$
where $f$ is frequency and $f_c$ is a constant (its position is shown in the Fig. 1 by a dotted arrow). Such spectrum is usual characteristic of smooth bounded deterministic dynamics (see, for instance, Refs. \cite{mm}-\cite{fm} and references therein). 

   With increase of the parameter $\delta$ the convection dynamics becomes more and more complicated and at the value $\delta \simeq 3$ time-series of the $N(t)$ fluctuations are already random-like (the dynamics in this case is dominated mainly by the the roll pinch-off events \cite{pcfg}). In order to understand this transition one should take into account that with the increase of $\delta$ the parameter $f_c$ in the Eq. (4) becomes fluctuating and one should use an ensemble average over this fluctuating parameter (with certain distribution $P(f_c)$) to compute the power spectrum 

$$
E(f) = \int P(f_c) ~\exp-(f/f_c)~ df_c \eqno{(5)}
$$  

To find $P(f_c)$ let us recall that despite the thermal convection is not an Hamiltonian system there is a possibility to introduce an effective action $I_e$ for such systems in the case of chaotic/stochastic dynamics (see Ref. \cite{E} and references therein). The basic dynamic relationship involving the action is 
$$
u_c \propto I_e^{1/2} f_c^{1/2}  \eqno{(6)}
$$
where $u_c$ is a characteristic velocity and $I_e$ can be considered as an adiabatic invariant \cite{suz}.
   Providing Gaussian (normal) distribution of the characteristic velocity (see, for instance, Ref. \cite{my}) we obtain 
$$
P(f_c) \propto f_c^{-1/2} \exp-(f_c/4f_{\beta})  \eqno{(7)}
$$
where $f_{\beta} =$ constant. 

   Substitution of the Eq. (7) into Eq. (5) results in
$$
E(f) \propto \exp-(f/f_{\beta})^{1/2}  \eqno{(8)}
$$
   
   Figure 2 shows a windowed power spectrum of the $N(t)$ fluctuations obtained at $\delta = 3$. The spectral data were taken from Fig. 3 of the Ref. \cite{pcfg}. The dashed curve in the Fig. 2 indicates the stretched exponential spectrum Eq. (8). It follows from comparison of the Figs. 1 and 2 that the stretched exponential power spectrum in the Fig. 2 is tuned to the low-frequency edge of the exponential spectrum shown in Fig. 1 - $f_{\beta}$ (the second dotted arrow). It is natural if the distributed chaos at $\delta = 3$ is a result of the development of the deterministic chaos at $\delta =  0.614$
\begin{figure} \vspace{-1.7cm}\centering
\epsfig{width=.45\textwidth,file=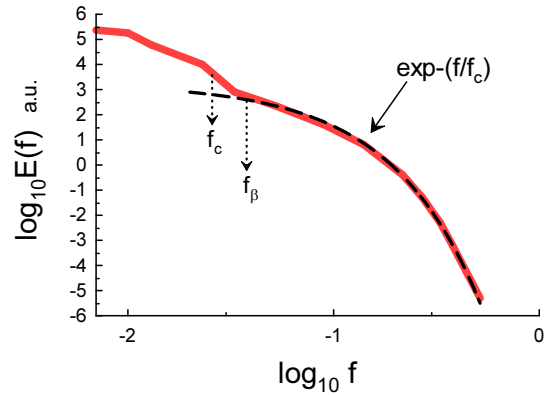} \vspace{-4.5cm}
\caption{Windowed power spectrum of the $N(t)$ fluctuations at $\delta =  0.614$.} 
\end{figure}
\begin{figure} \vspace{-0.3cm}\centering
\epsfig{width=.45\textwidth,file=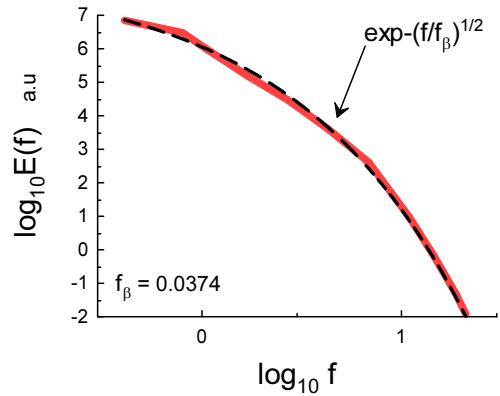} \vspace{-4.4cm}
\caption{Windowed power spectrum of the $N(t)$ fluctuations at $\delta =  3$.} 
\end{figure}

\section{Kolmogorov-Bolgiano-Obukhov distributed chaos}
  
\subsection{Inertial-Buoyancy Invariant}

   It can be readily shown that the system Eqs. (1-3) has an ideal (at $\nu=\kappa=0$) invariant \cite{kcv}
$$
\mathcal{E} = \int_V ({\bf u}^2 -S\sigma g \frac{H}{\Delta}\theta^2) ~ d{\bf r}   \eqno{(9)}
$$    
$V$ is the spatial volume.

\begin{figure} \vspace{-1.2cm}\centering
\epsfig{width=.45\textwidth,file=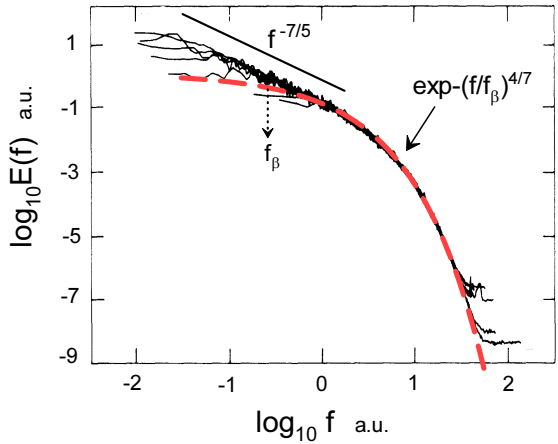} \vspace{-3.7cm}
\caption{Power spectrum of the temperature fluctuations in the bulk of the Rayleigh-B\'{e}nard convection in helium gas at $Ra$ in the range from $7 \times 10^6$ to $7.3 \times 10^{10}$.} 
\end{figure}
\begin{figure} \vspace{-0.3cm}\centering
\epsfig{width=.45\textwidth,file=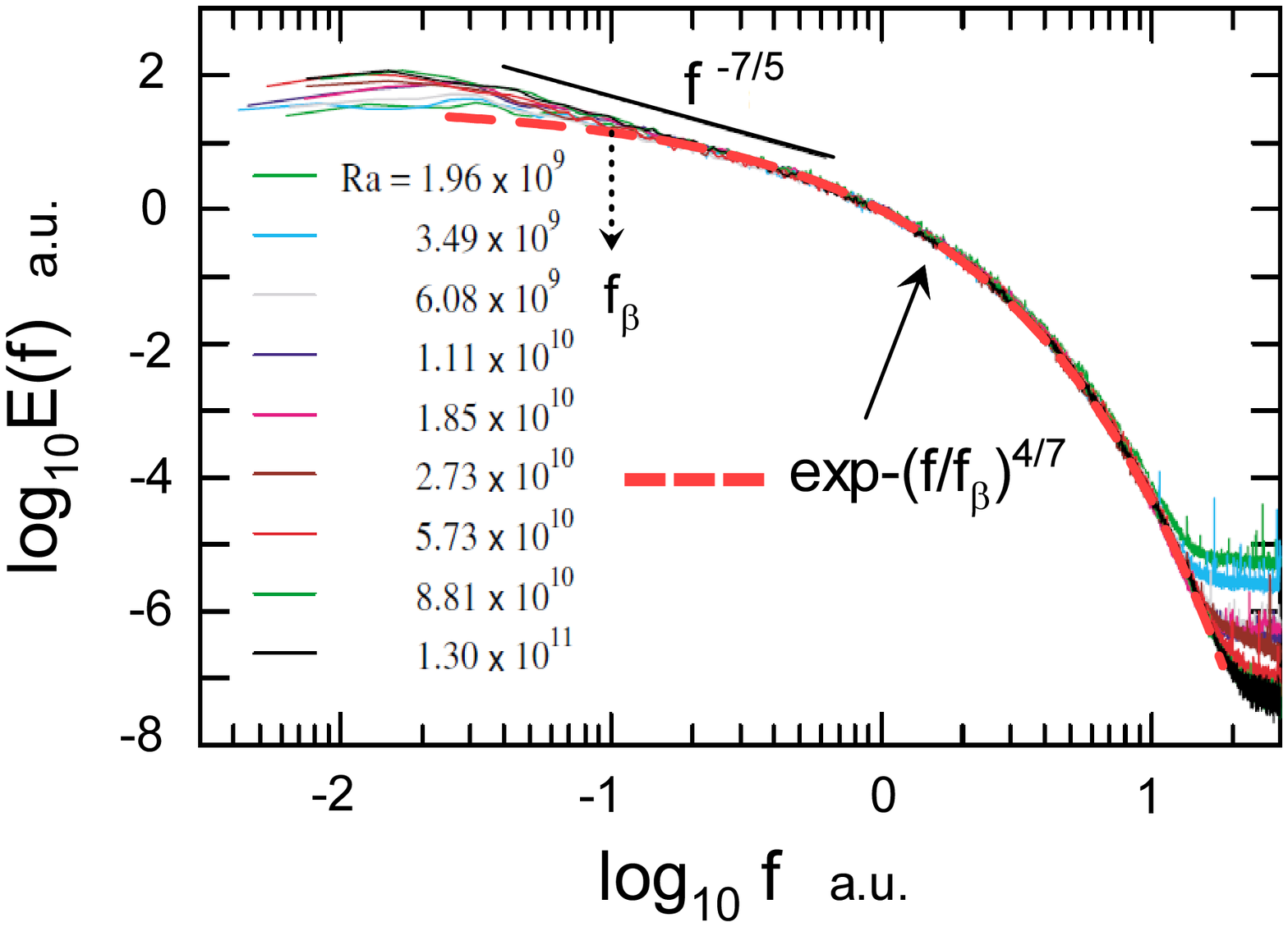} \vspace{-4.1cm}
\caption{Power spectrum of the temperature fluctuations in the bulk of the Rayleigh-B\'{e}nard convection in water at $Ra$ in the range from $4.1 \times 10^8$ to $1.3 \times 10^{11}$.} 
\end{figure}

 It is a generalization of the energy invariant of the Navier-Stokes system. Then, using corresponding generalization of the Kolmogorov-Bolgiano-Obukhov phenomenology on the inertial-buoyancy range of scales we can obtain from the dimensional consideration relationship between characteristic temperature fluctuation $\theta_c$ and characteristic wavenumber scale $k_c$
$$
\theta_c\propto  (\sigma g)^{-1} \varepsilon^{2/3} k_c^{1/3} \eqno{(10)}
$$
where the generalized dissipation rate 
$$
\varepsilon = \left|\frac{d\langle{\bf u}^2 -S\sigma g \frac{d}{\Delta}\theta^2 \rangle}{dt}\right| \eqno{(11)}
$$
here $\langle ... \rangle$ is spatial averaging.

  For stable stratification $S=-1$ and when the buoyancy forces dominate over the inertial forces (the Bolgiano-Obukhov case \cite{my}) one can use 
$$
\varepsilon_b = \left|\frac{d\langle \theta^2 \rangle}{dt}\right| \eqno{(12)}
$$  
instead of the $\varepsilon$ and then one obtains
$$
\theta_c \propto (\sigma g)^{-1/5} \varepsilon_b^{2/5} k_c^{1/5}  \eqno{(13)}
$$
instead of the relationship Eq. (10).\\

 For the Rayleigh-B\'{e}nard convection $S=1$ and situation is more complex (although, formally, one can apply the above consideration to this case as well). In the Refs. \cite{pz},\cite{lvov},\cite{fl} some elaborated arguments were given regarding applicability of the Bolgiano-Obukhov approach to the Rayleigh-B\'{e}nard convection. In particular, it was argued that the spatial scaling power spectrum of the temperature fluctuations in the buoyancy dominated range of scales is 
$$
E(k) \propto (\sigma g)^{-2/5} \varepsilon_b^{4/5} k^{-7/5} \eqno{(14)}
$$    
similar to stable stratification. The same arguments can be applied for applicability of the Eq. (13) for the Rayleigh-B\'{e}nard convection as well.\\

\subsection{Experiments - I}

   Figures 3-5 show power spectra obtained in laboratory experiments reported in the Refs. \cite{wu},\cite{zx},\cite{nssd}. The Taylor hypothesis directly relates locally measured frequency spectra to the spatial (wavenumber) ones \cite{my}. Therefore, the solid straight lines are drawn in the Figs. 3-5 to indicate correspondence to the spectrum Eq. (14) (see also Ref. \cite{b1}).  \\
   
   The spectral data for Fig. 3 were taken from the Fig. 2 of the Ref. \cite{wu} (the spectrum was normalized, see the Ref. \cite{wu} for the details) and were obtained in a cylindrical cell with the aspect ratio $\Gamma =0.5$ filled with helium gas. The measurements were
made in the center of the cell. Therefore they can be regarded as the bulk measurements. The Rayleigh number $Ra$ was in the range from $7 \times 10^6$ to $7.3 \times 10^{10}$.  \\
   
   The spectral data for Fig. 4 were taken from the Fig. 1a of the Ref. \cite{zx} and were obtained in water in two cylindrical convection cells with the aspect rations $\Gamma =1$ and 0.5 (for the first cell $Ra$ was in the range $4.1 \times 10^8$ to $1.85 \times 10^{10}$, whereas for the second cell in the range $2.7 \times 10^{10}$ to  $1.3 \times 10^{11}$, corresponding value of the Prandtl number $Pr = 5.3$ and $4.3$). The measurements were
made by traversing the thermistor from the midheight of the cell to its bottom, along central axis of the cell. Therefore they can be regarded as the bulk measurements. \\

  The measurements of the spectral data used in the Fig. 5 were made in cryogenic helium gas near a side wall (outside the boundary layer) at the midheight of a cylindrical cell ($\Gamma =1/2$, $Ra = 1.5 \times 10^{11}$ and $Pr \sim 1$). \\

\subsection{Spatial distributed chaos}

  The exponential power spectra are typical for deterministic chaos (or for onset of turbulence) not only in the frequency domain but for the wavenumber domain as well (see, for instance Refs. \cite{mm},\cite{kds} and references therein)
$$ 
E(k) \propto \exp(-k/k_c)  \eqno{(15)}
$$

  It is known that at $Pr \sim 1$ transition to turbulence occurs at $Ra \sim 10^6$ (see, for instance, Ref. \cite{v}). Figure 6 shows spatial (wavenumber) power spectrum of temperature fluctuations at $Ra = 6.6 \times 10^6$ and $Pr =1$. The spectrum was obtained in a direct numerical simulation (with free-slip boundary conditions) reported in Ref. \cite{mv} and the spectral data shown in the Fig. 6 were taken from the Fig. 11 of the Ref. \cite{mv}. The spectrum has two branches, but the upper branch contains rather small number of the Fourier modes and the only lower branch seems to be physically significant. The dashed curve is drawn to indicate correspondence to the exponential Eq. (15). \\
  
  With transition to turbulence, analogously to the temporal description, fluctuations of the characteristic wavenumber $k_c$ can be taken into account by the ensemble averaging 
$$
E(k) \propto \int_0^{\infty} P(k_c) \exp -(k/k_c)dk_c \propto \exp-(k/k_{\beta})^{\beta} \eqno{(16)}
$$  
where the stretched exponential in the right-hand side of the Eq. (16) is a generalization of the Eq. (8). In this case    
the probability distribution $P(k_c)$ can be estimated for large $k_c$ from the Eq. (16) \cite{jon}
$$
P(k_c) \propto k_c^{-1 + \beta/[2(1-\beta)]}~\exp(-\gamma k_c^{\beta/(1-\beta)}) \eqno{(17)}
$$     
where $\gamma$ is a constant.

 The scaling relationships  Eqs. (10) and (13) can be written in a general form as   
$$
\theta_c \propto  k_c^{\alpha}   \eqno{(18)}
$$

 If $\theta_c$ has Gaussian distribution (with zero mean) \cite{my} a relationship between $\alpha$ and $\beta$ 
$$
\beta = \frac{2\alpha}{1+2\alpha}  \eqno{(19)}
$$
follows immediately from the Eqs. (17-18). \\

  For the Eq. (10) the $\alpha = 1/3$ and we obtain from the Eq. (19) $\beta = 2/5$, i.e. for the {\bf inertial-buoyancy} regime
$$
E(k) \propto \exp-(k/k_{\beta})^{2/5},  \eqno{(20)}
$$
whereas for the Eqs. (13) the $\alpha = 1/5$ and we obtain from the Eq. (19) $\beta = 2/7$, i.e. for the {\bf buoyancy dominated} regime
$$
E(k) \propto \exp-(k/k_{\beta})^{2/7}  \eqno{(21)}
$$
   
\begin{figure} \vspace{-1.6cm}\centering
\epsfig{width=.45\textwidth,file=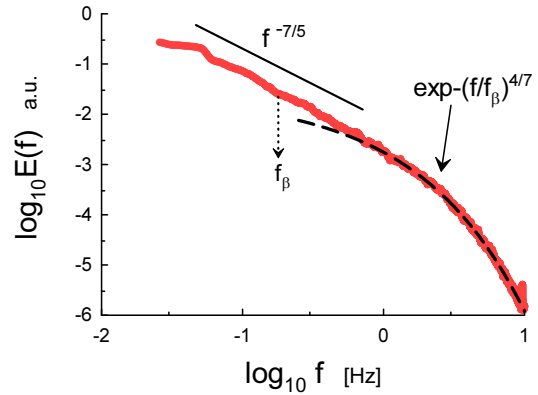} \vspace{-4.2cm}
\caption{Power spectrum of the temperature fluctuations near the side wall for the Rayleigh-B\'{e}nard  convection at the midheight of a cylindrical cell with $Ra = 1.5 \times 10^{11}$ and $Pr \sim 1$.} 
\end{figure}
  
\begin{figure} \vspace{-0.3cm}\centering
\epsfig{width=.45\textwidth,file=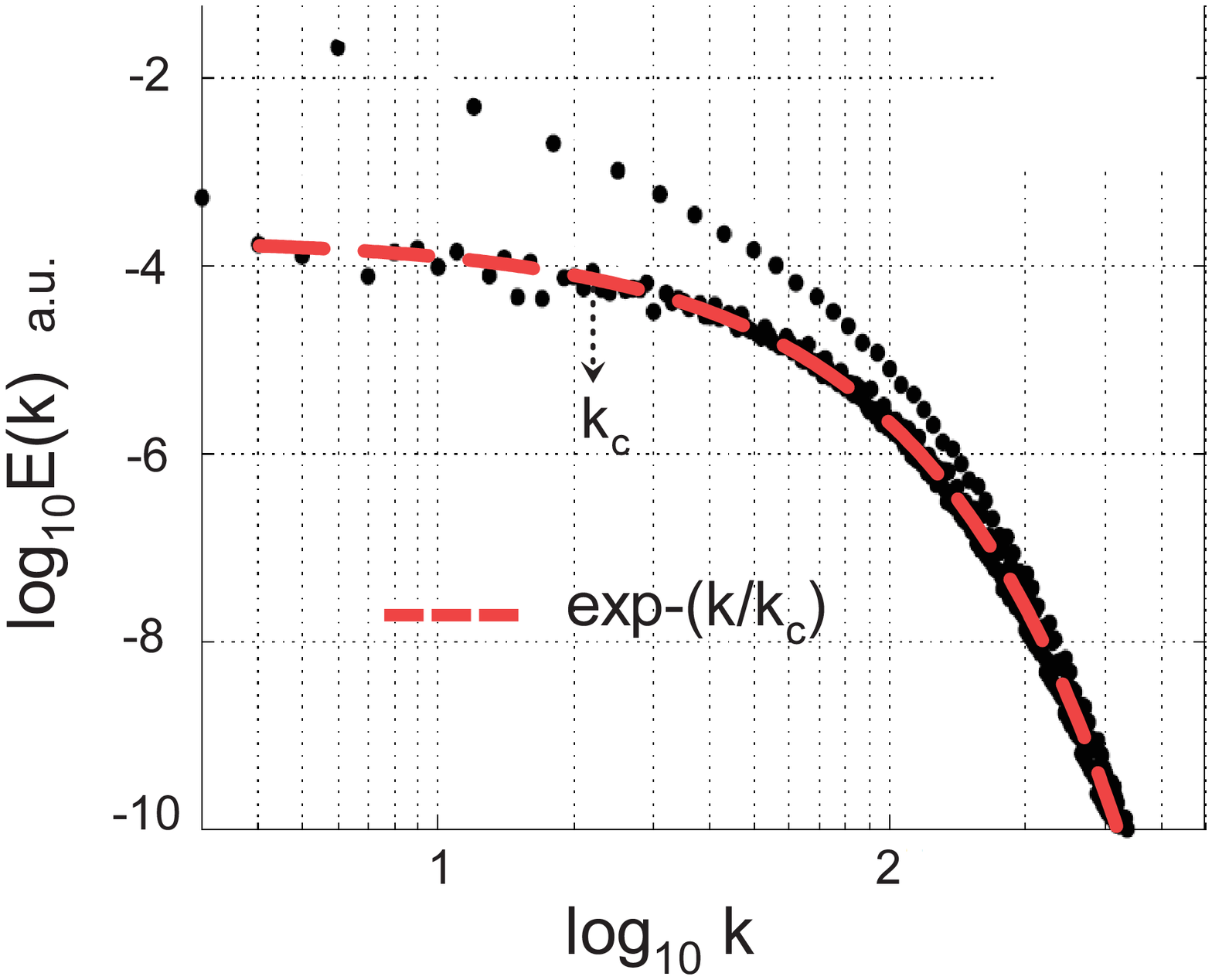} \vspace{-3.33cm}
\caption{Power spectrum of the temperature fluctuations for the Rayleigh-B\'{e}nard  convection at $Ra = 6.6 \times 10^{6}$ and $Pr = 1$. } 
\end{figure}
\begin{figure} \vspace{-1.9cm}\centering
\epsfig{width=.45\textwidth,file=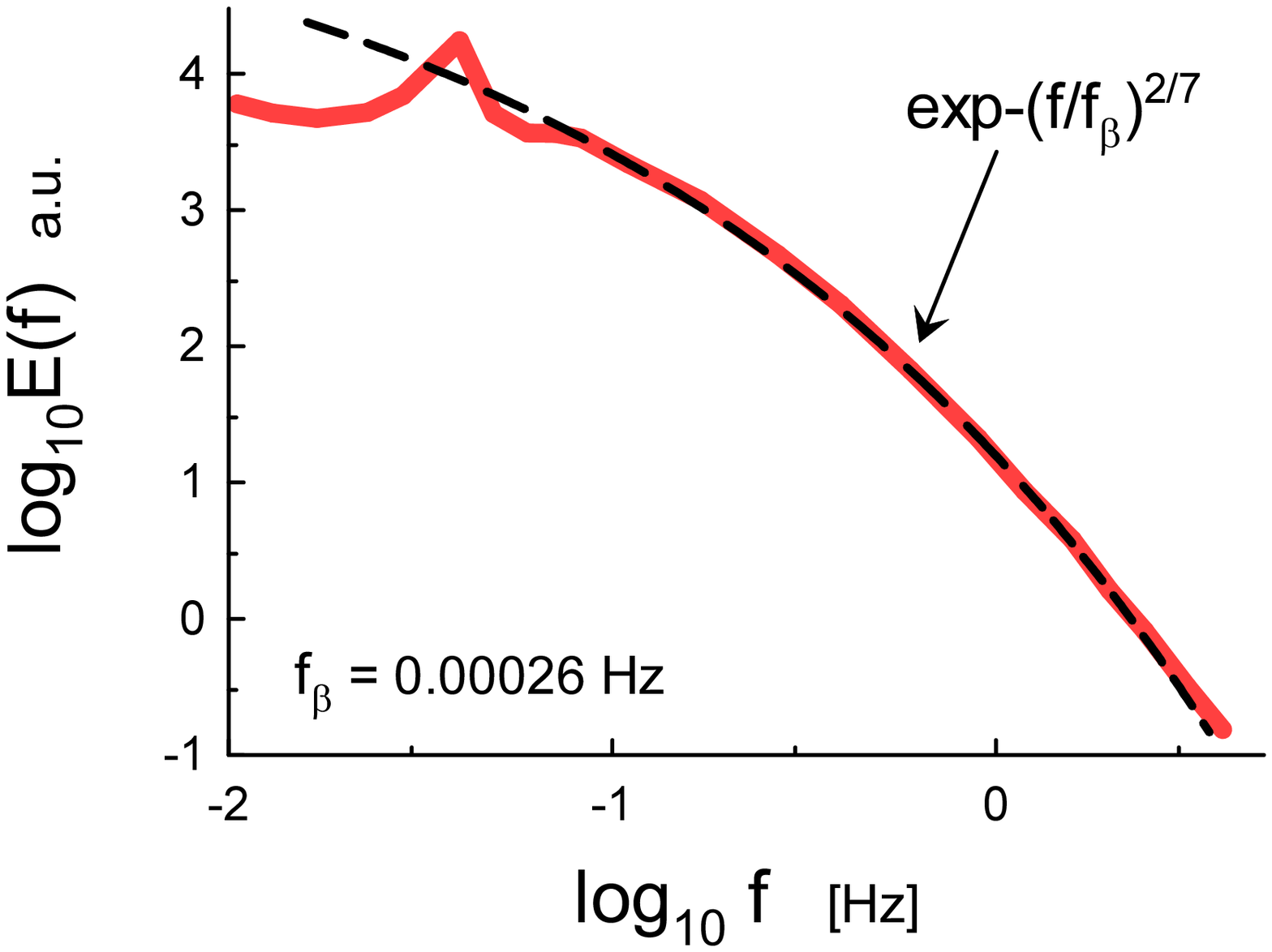} \vspace{-3.97cm}
\caption{Power spectrum of the temperature fluctuations in the bottom corner of a cylindrical cell for the Rayleigh-B\'{e}nard  convection at $Ra = 2.05 \times 10^{11}$ and $Pr = 0.8$. } 
\end{figure}
\begin{figure} \vspace{-0.3cm}\centering
\epsfig{width=.45\textwidth,file=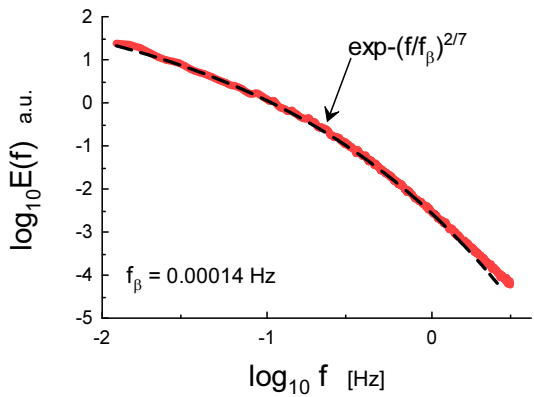} \vspace{-4.2cm}
\caption{Power spectrum of the temperature fluctuations near the side wall for the Rayleigh-B\'{e}nard  convection at the midheight of a cylindrical cell with $Ra = 1.63 \times 10^{13}$ and $Pr = 0.8$.} 
\end{figure}

\subsection{Experiments - II}

  We will consider the distributed chaos for the inertial-buoyancy regime  in more detail in Section V and here let us consider several observations of the distributed chaos for the buoyancy dominated regime. \\
  
   In Ref. \cite{he1} a laboratory experiment with the Rayleigh-B\'{e}nard  convection was performed in a cylindrical cell with aspect ratio $\Gamma =1$ at $Ra = 2.05 \times 10^{11}$ and $Pr = 0.8$. Position of the thermistor was $z/H=0.019$ and 
$(R-r)/R= 0.064$ (where $R$ is radius of of the cell). It is the bottom corner of the cylindrical cell and it should be noted that for this position $z/\lambda_{\theta} \simeq 12.7$, where $\lambda_{\theta}$ is the thickness of a thermal sublayer adjacent to the bottom plate where the heat flux is mostly due to the thermal diffusion.  \\

   Figure 7 shows power spectrum of the temperature fluctuations measured at the above mentioned position (the spectral data were taken from Fig. 2 of the Ref. \cite{he1}). The dashed curve in the Fig. 7 indicates the stretched exponential spectrum Eq. (21).\\
   
   In Ref. \cite{he2} a laboratory experiment with the Rayleigh-B\'{e}nard  convection was performed in a cylindrical cell with aspect ratio $\Gamma =0.5$ at $Ra = 1.63 \times 10^{13}$ and $Pr = 0.8$. Position of the thermistor was $z/H=0.4933$ and  $(R-r)/R= 0.0356$ (where $R$ is radius of of the cell), i.e. near the side-wall and approximately at midheight of the cell.   The thermistor position is rather close to that used in the second experiment described in the subsection {\bf B}, also value of $Pr$ is about the same only the value of $Ra$ is different. However, this difference in the value of $Ra$ is crucial because the value $Ra = 1.63 \times 10^{13}$ is the {\it critical} one at the transition from the so-called classical to ultimate regime \cite{he2}. \\
    
    Figure 8 shows power spectrum of the temperature fluctuations measured at the above mentioned position. The dashed curve in the Fig. 8 indicates the stretched exponential spectrum Eq. (21). Comparison with the Fig. 4, corresponding to the classical regime, indicates that the buoyancy dominated regime at the large scales before the transition (the Eq. (14)) is considerably expanded at the critical (transitional) value of the $Ra \sim 10^{13}$ and the scaling Eq. (14) is replaced by the stretched exponential Eq. (21) (the distributed chaos).

\section{Buoyancy-helical distributed chaos}

   The inertial-buoyancy invariant $\mathcal{E}$ Eq. (9) is not sole inviscid invariant for the Rayleigh-B\'{e}nard  convection. Let us consider equation for helicity $h={\bf u}\cdot {\boldsymbol \omega}$ (here the vorticity ${\boldsymbol \omega} = \nabla \times {\bf u}$) corresponding to the Eq. (1) with $\nu =0$
$$
\frac{d\langle h \rangle}{dt}  = 2\sigma g e_z\langle \omega_z \theta \rangle \eqno{(22)} 
$$ 
  It follows from the Eq. (22) that the helicity generally is not an ideal invariant of the Rayleigh-B\'{e}nard  convection. However, one can expect that the correlation $\langle \omega_z \theta \rangle$ being considerable at large scales (on the coherent structures) is quickly decreased in the chaotic/turbulent motion when the spatial scales become smaller. Therefore, the second order moment of the helicity distribution (the Levich-Tsinober invariant of the Euler equation \cite{lt}) can be still considered as an ideal invariant of the Rayleigh-B\'{e}nard  convection. Indeed, let us divide the volume of motion into the cells $V_j$ subject to the boundary conditions ${\boldsymbol \omega} \cdot {\bf n}=0$ on the bounding surfaces of the cells - $S_j$, moving with the fluid \cite{lt},\cite{mt}. Then for the sells (where the spatial scales are small enough) the helicity, averaged over the cell, can be approximately considered as an ideal invariant. The second order moment can be defined as \cite{mt}
$$
I = \lim_{V \rightarrow  \infty} \frac{1}{V} \sum_j H_{j}^2  \eqno{(24)}
$$
where 
$$
H_j = \int_{V_j} h({\bf r},t) ~ d{\bf r}.  \eqno{(25)}
$$
 The sum of the invariants $H_j$ is an invariant.\\
 
In order to apply the Kolmogorov-Bolgiano-Obukhov phenomenology we should take into account that unlike $\mathcal{E}$ Eq. (9), which is quadratic invariant, the  $I$ Eq. (24) is a quartic invariant. Therefore, we should use $\varepsilon_I =|dI^{1/2}/dt|$ in the Eq. (10) instead of the $\varepsilon$
$$
\theta_c\propto  (\sigma g)^{-1} \varepsilon_I^{2/3} k_c^{2/3} \eqno{(26)}
$$
Then $\alpha =2/3$ and $\beta =4/7$ (the Eqs. (18-19)), i.e.
$$
E(k) \propto \exp-(k/k_{\beta})^{4/7}  \eqno{(27)}
$$
in this case. \\ 
\begin{figure} \vspace{-1.55cm}\centering
\epsfig{width=.45\textwidth,file=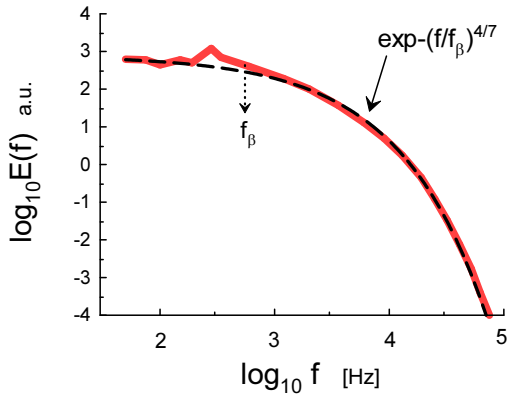} \vspace{-4.3cm}
\caption{Power spectrum of the temperature fluctuations in the bulk of the Rayleigh-B\'{e}nard convection in helium gas at $Ra=1.1\times 10^8$ and $Pr \simeq 1$ .} 
\end{figure}
\begin{figure} \vspace{-0.3cm}\centering
\epsfig{width=.45\textwidth,file=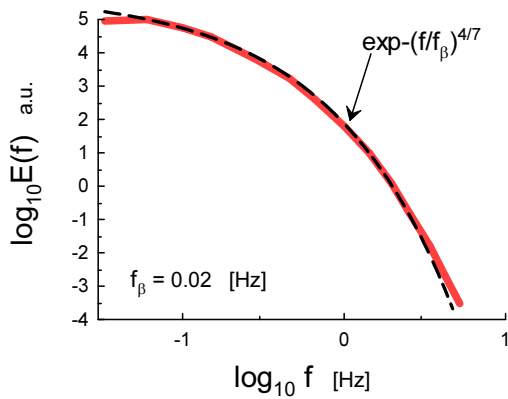} \vspace{-4.37cm}
\caption{Power spectrum of the temperature fluctuations near bottom of the Rayleigh-B\'{e}nard convection in water at $Ra=4.07 \times 10^9$ and $Pr \simeq 7$.} 
\end{figure}
\begin{figure} \vspace{-1.6cm}\centering
\epsfig{width=.45\textwidth,file=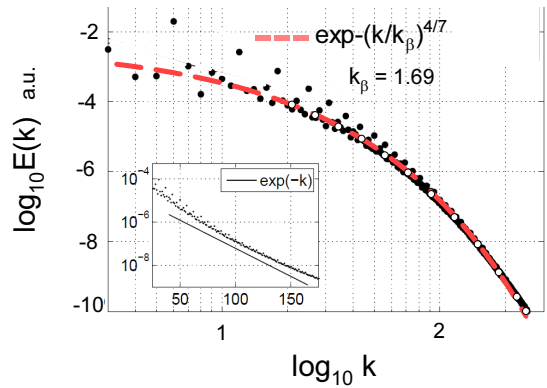} \vspace{-4.27cm}
\caption{Power spectrum of the temperature fluctuations for the Rayleigh-B\'{e}nard  convection at $Ra = 2.6 \times 10^{6}$ and $Pr = 0.02$.} 
\end{figure}
  
  The stretched exponential was for the first time suggested (empirically) to fit the spectra for the Rayleigh-B\'{e}nard  convection in the Ref. \cite{wu} (see Fig. 3 and corresponding description in the previous Section). The dashed curve in this figure indicates the stretched exponential Eq. (27). The dotted arrow is drawn to indicate position of the $f_{\beta}$. Analogous situation can be observed in Figs. 4 and 5.\\ 
  
  Figure 9 shows an individual (non-normalized) spectrum taken from those shown in the Fig. 3 and corresponding to comparatively small value of the Rayleigh number $Ra = 1.1 \times 10^8$. Let us recall that these spectra were obtained approximately at the center of a cylindrical cell (bulk spectra).  The dashed curve in this figure indicates the stretched exponential Eq. (27). The dotted arrow is drawn to indicate position of the $f_{\beta}$.\\
  
  Figure 10 shows experimentally obtained power spectrum of the temperature fluctuations near bottom of a vertical cylindrical cell ($\Gamma \simeq 1$) with the Rayleigh-B\'{e}nard convection at $Ra=4.07 \times 10^9$ and $Pr \simeq 7$. The spectral data were taken from Fig. 1 of the Ref. \cite{xz}.  The dashed curve in this figure indicates the stretched exponential Eq. (27).  \\
  
  It is known that for small Prandtl numbers effective Rayleigh number $Ra$ is usually much larger than for $Pr \sim 1$. Figure 11 shows power spectrum of the temperature fluctuations computed in the same direct numerical simulation as that shown in the Fig. 6 but now for $Pr \simeq 0.02$ and $Ra = 2.6 \times 10^6$. The spectral data were taken from Fig. 5 of the Ref. \cite{mv}.  The dashed curve in this figure indicates the stretched exponential Eq. (27). The insert in this figure shows the spectrum in the linear-log scales for comparison with a pure exponential - the straight line in the insert.  It should be noted that this global wavenumber spectrum is obviously determined by the bulk of the flow.

\section{Transition to the ultimate state}
\begin{figure} \vspace{-1.6cm}\centering
\epsfig{width=.45\textwidth,file=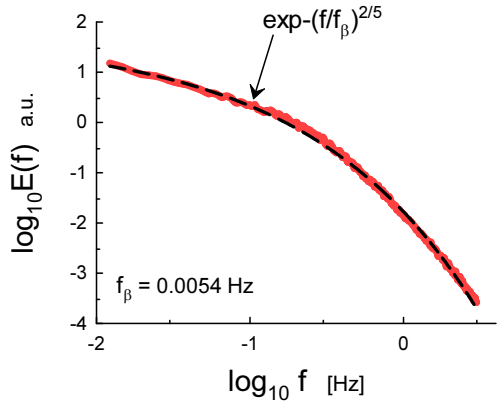} \vspace{-3.97cm}
\caption{Power spectrum of the temperature fluctuations in the bottom corner of a cylindrical cell for the Rayleigh-B\'{e}nard  convection at $Ra = 1.63 \times 10^{13}$ and $Pr = 0.8$. } 
\end{figure}
\begin{figure} \vspace{-0.5cm}\centering
\epsfig{width=.45\textwidth,file=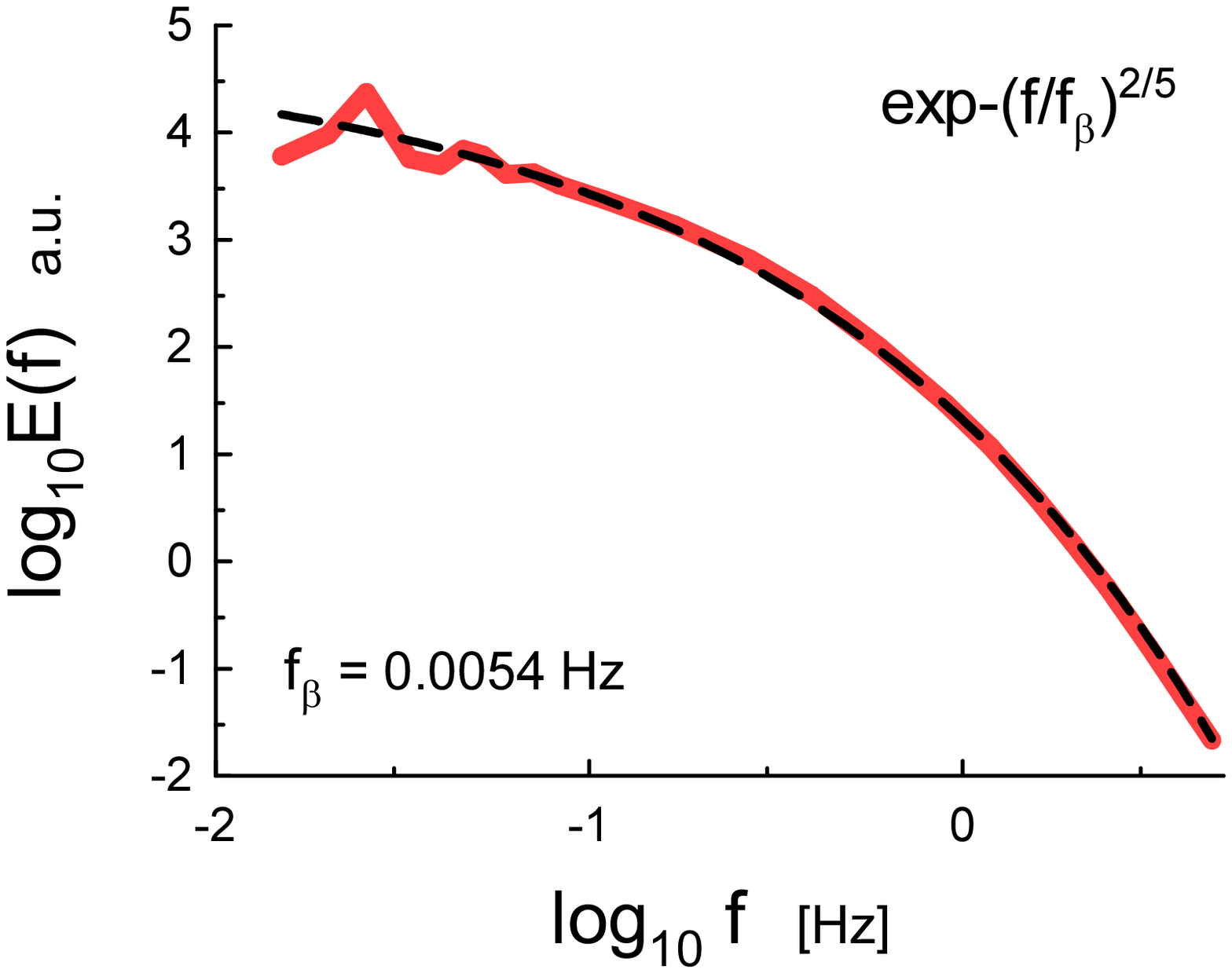} \vspace{-4.7cm}
\caption{Power spectrum of the temperature fluctuations near the side wall for the Rayleigh-B\'{e}nard  convection at the midheight of a cylindrical cell with $Ra = 1.35 \times 10^{14}$ and $Pr = 0.8$.} 
\end{figure}
\begin{figure} \vspace{-1.2cm}\centering
\epsfig{width=.45\textwidth,file=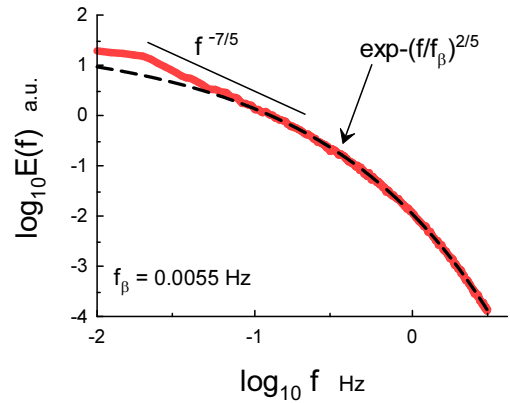} \vspace{-4.37cm}
\caption{As in Fig. 13 but at $Ra = 1.08 \times 10^{15}$.} 
\end{figure}
    Transition from so-called classical to ultimate state in the Rayleigh-B\'{e}nard  convection has been discussed for a long time (see for instance Refs. \cite{gl},\cite{do} and references therein)) but the conclusions are still elusive. It is experimentally shown in the Refs. \cite{he2},\cite{he3} that for $Pr \sim 1$ in the range from $Ra \simeq 10^{13}$ to $Ra \sim 10^{14}$ there is a transition in character of the heat transport produced by the Rayleigh-B\'{e}nard convection that can be related to the expected transition from the classical to ultimate state. Let us look at this problem using the above described approach.\\
    
    We have already seen that the temperature spectrum near the side wall at midheight of the cylindrical cell has been changed at $Ra \sim 10^{13}$ (cf. Fig. 5 and 8). However, the both spectra Eq. (21) and Eq. (27) belong to the classical state and the change in the spectrum at this location will be continued with further increase of $Ra$ (see with Figs. 13 and 14). Therefore, it is more instructive to look at the change in spectrum which occurs at the $Ra \sim 10^{13}$ in the bottom corner of the cylindrical cell (where the main generation of the plums is expected). Figure 12 shows power spectrum of the temperature fluctuations at the bottom corner of a cylindrical cell for the Rayleigh-B\'{e}nard  convection at $Ra = 1.63 \times 10^{13}$ and $Pr = 0.8$. The spectral data were obtained in the same experiment \cite{he2} as those shown in the Fig. 8 but for $z/H = 0.0179$ (i.e. in the bottom corner). The dashed curve is drawn in the figure to indicate correspondence to the inertial-buoyancy regime Eq. (20) (cf. Fig. 7 corresponding to the buoyancy dominated regime Eq. (21) at $Ra = 2.05 \times 10^{11}$, i.e. at the classical state). This change of the regimes from the buoyancy dominated to the inertial-buoyancy one can be considered as a crucial since the inertial-buoyancy regime does not appear at the classical state.\\
    
    Moreover the spectrum Eq. (20) corresponding to the inertial-buoyancy regime replaces also the buoyancy spectrum Eq. (21) near the side wall (at the midheight of the cell) at further increase of $Ra$. Figure 13 shows power spectrum of the temperature fluctuations near the side-wall for the Rayleigh-B\'{e}nard convection at the midheight of a cylindrical cell with $Ra = 1.35 \times 10^{14}$ and $Pr = 0.8$ (compare with the Fig. 8). The spectral data were taken from Fig. 2 of the Ref. \cite{he1}. The inertial-buoyancy spectrum Eq. (20) can be also observed at the midheight for $Ra = 1.08 \times 10^{15}$ (see Fig. 14, compare also the Fig. 14 with the Fig. 5 corresponding to the classical state) and in the bottom corner of the cell as well (see Fig. 15). The spectral data for the Figs. 14 and 15 were obtained in the experiment reported in the Ref. \cite{he2}. It also should be noted that the value of $f_{\beta}$ (i.e., due to the Taylor hypothesis, of the $k_{\beta}$) is the same at the situations shown in Figs. (12-14) (the situation corresponding to the Fig. 12 is the transitional one).\\
    
    Thus the inertial-buoyancy regime Eq. (20) replaces the classical regimes Eq. (21) and Eq. (27) at the ultimate state.

\section{Discussion and Conclusions}

\begin{figure} \vspace{-1.3cm}\centering
\epsfig{width=.45\textwidth,file=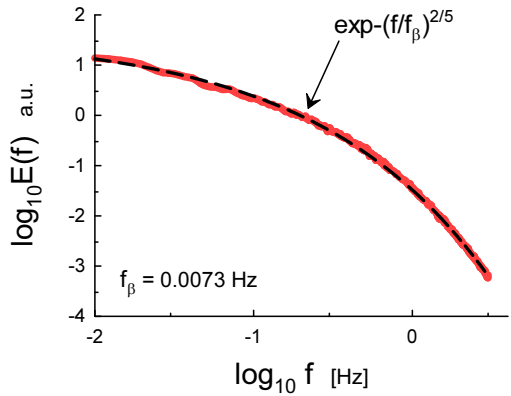} \vspace{-4.6cm}
\caption{As in Fig. 12 but at $Ra = 1.08 \times 10^{15}$.} 
\end{figure}

 At sufficiently large Rayleigh number $Ra$ the deterministic chaos becomes unstable with fluctuating parameters. The unstable (fluctuating) deterministic chaos can be described in the terms of the distributed chaos with a stretched exponential spectrum. This process is controlled by chaotic/stochastic action (a generalization of the Hamiltonian action \cite{E}).  Further increase in $Ra$ results in appearance of different regimes of the distributed chaos (turbulence). These regimes are dominated by the ideal invariants (with $\nu = \kappa =0$) of the Rayleigh-B\'{e}nard convection Eqs. (1-3), which can be considered as adiabatic invariants for the extended inertial range of scales for the dissipative (diffusive) convection.
 
  Two distributed chaos regimes with temperature fluctuation power spectra given by the Eqs. (21) and Eq. (27) are typical for the classical state of the Rayleigh-B\'{e}nard  convection - the first is applicable for situations where the buoyancy forces dominate the inertial ones (the generalized Bolgiano-Obukhov approximation) and second is applicable where the helicity effects are dominating (the Levich-Tsinober invariant approximation). Naturally, the buoyancy approximation is more appropriate at the bottom corner where the plumes mainly originate, whereas the helicity based approximation is more appropriate in the bulk of the flow. 
  
   At transition to the ultimate state the classical regimes are replaced by the inertial-buoyancy regime Eq. (20) (although the Bolgiano-Obukhov scaling Eq. (14) can still appear at the large scales - Fig. 14).\\

\section{Acknowledgement}

I thank E. Levich for stimulating discussions, and X. He, J.J. Niemela and K.R. Sreenivasan for sharing their data.

\end{document}